\documentclass[aps,showpacs,amsmath,amssymb]{revtex4}

\usepackage{graphicx,epsf}
\usepackage{dsfont}
\usepackage[]{latexsym}
\newcommand{\be}{\begin{eqnarray}}
\newcommand{\ee}{\end{eqnarray}}

\begin{document}
\title{Zero point energy on extra dimensions: Noncommutative Torus}
\author{S. Fabi}
\email{fabi001@bama.ua.edu}
\affiliation{Department of Physics and Astronomy, The University of
Alabama, Box 870324, Tuscaloosa, AL 35487-0324, USA}
\author{B. Harms}
\email{bharms@bama.ua.edu}
\affiliation{Department of Physics and Astronomy, The University
of Alabama, Box 870324, Tuscaloosa, AL 35487-0324, USA}
\author{G. Karatheodoris}
\email{karat002@bama.ua.edu}
\affiliation{Department of Physics and Astronomy, The University
of Alabama, Box 870324, Tuscaloosa, AL 35487-0324, USA}
\begin{abstract}
In this paper we calculate the zero point energy density experienced
by observers on $\mathcal{M}^4$ due to a massless scalar field
defined throughout $\mathcal{M}^4 \times \mathbf{T}^2_{F}$, where
$\mathbf{T}^2_{F}$ are fuzzy extra dimensions. Using the Green's
function approach we calculate the energy density for the
commutative torus and the fuzzy torus. We calculate then the energy
density for the fuzzy torus using the Hamiltonian approach.
Agreement is shown between Green's function and Hamiltonian
approaches.

\end{abstract}
\maketitle
\large
\section{Introduction}\label{intro}

The zero-point energy of the Universe remains one of the
fundamental mysteries of physics.  The Universe is known to be
accelerating, and so far the cause of this phenomenon (called dark
energy) has defied explanation in terms of conventional classical
or quantum physics. Since the Universe's zero-point energy is one
the characteristics determining its fate, it is important to
determine the factors which may contribute to the value of the
zero-point energy.  A possible source of vacuum energy is the
Casimir effect. The contribution of this effect to the vacuum
energy depends upon the topology of spacetime, the number of
spacetime dimensions, the type of field existing in the spacetime
and possibly other phenomena, such as the commutativity or
noncommutativity of spacetime.  The present paper is part of a
series of attempts to determine under which conditions the effects
described above can contribute to the vacuum energy density of the
Universe.
\par
In a previous paper \cite{fabi} we explored the possibility that
dark energy is a manifestation of the Casimir energy from extra
dimensions with the topology of a noncommutative $\mathbf{S}^2$. We
found that the value of the energy density present on $\mathcal{M}^4
\times \mathbf{S}_{F}^{2}$ is positive, i.e. it provides dark
energy, and we calculated the radius of $\mathbf{S}_{F}^{2}$ for a
chosen value of the size of the representation of the noncommutative
algebra. An exciting approach to noncommutative extra dimensions, in
which the said dimensions are even \emph{dynamically} generated (and
are spheres) is described in \cite{aschieri}.
\par The purpose of this paper is to repeat the calculation with
the extra dimensions being the fuzzy torus $\mathbf{T}_{F}^2$
instead of $\mathbf{S}_{F}^2$. In \cite{gomis},\cite{nam} the study
of a self-coupled massive scalar field on $ \mathcal{M}^{1,D} \times
\mathbf{T}_{\theta}^{2}$ has been considered. We consider a massless
scalar field. We take two different approaches to the problem: the
first is based on the Green's function method used by R. Kantowski
and K. Milton in~\cite{milton1}, and the second is based on the
Hamiltonian approach, as done in ~\cite{jabbari}. With the use of
the Green's function we perform the calculations for both the
commutative and fuzzy cases: $\mathbf{T}_{F}^{2}$ which has finite
dimensional representations. In \cite{jabbari}, using the
Hamiltonian approach, the energy density was calculated for
$\mathds{R}\times \mathbf{T}_{F}^{2}$; we generalize the manifold to
the $\mathcal{M}^4 \times \mathbf{T}_{F}^{2}$ case using the same
approach. In the case of the fuzzy torus, the energy density
calculated using the Green's function agrees with that calculated
using the Hamiltonian formalism.
\par Regardless of the method, the energy density present on
$\mathcal{M}^4$ due to the extra dimensions with topology of
$\mathbf{T}_{F}^{2}$ turns out to be negative for each possible
value of the size of the representation $N$ of the algebra for the
torus and therefore it cannot be considered as a source for dark
energy. The energy density we obtain with the two different
approaches has the same form as obtained in the case
$\mathcal{M}^4\times \mathbf{S}_{F}^{2}$ \cite{fabi}
 \be\label{U}
  u_{\rho}=\frac{\alpha}{\rho^{4}}\ln\frac{\rho}{b}
 \label{uc1}
 \ee
 where  $\rho$ is the radius of the torus (the two radii are chosen to be the same), $b$ a momentum cutoff
  and $\alpha$ is a negative constant. In the fuzzy case the value of $\alpha$ turns out to be
  the same using either the Green's function or the Hamiltonian approach.
We also calculate the numerical value of the energy density in the
case of maximum noncommutativity $N=2$

\section{The Noncommutative Torus }
The noncommutative 2-torus is defined in terms of operators
$\hat{U}_{i}~(i=1,2)$ satisfying
 \be\label{comm}
\hat{U}_{i}\hat{X}_{j}\hat{U}_{i}^{-1}&=&X_{j}+\delta_{ij} 2\pi R_{j}\textbf{1}\\
 U_{i}U_{j}&=&e^{2\pi \theta_{ij}}U_{j}U_{i}
 \ee
where $X_{i}$ are the coordinate operators, $R_{i}$ the
compactification radii (set both equal to $\rho$) and the
 dimensionless noncommutativity parameter $\theta_{ij}=\theta \epsilon_{ij}$.
 From the definition above it follows that
 \be
 \hat{U}_{i}=e^{i\hat{\sigma}_{i}},\qquad \hat{\sigma}_{i}=\frac{\hat{X}_{i}}{\rho}
 \ee
 where the $\hat{\sigma}_{i}$ are dimensionless and satisfy
 \be
 [\hat{\sigma}_{i},\hat{\sigma}_{j}]=2\pi\theta\epsilon_{ij}.
 \ee
The $\lim \theta \rightarrow 0$ (the commutative limit) gives
$[\hat{X}_{i},\hat{X}_{j}]=0$. The definition of $\hat{U}_{i}$
allows a function defined on the $T_{\theta}^{2}$ to be expanded in
an operator Fourier series \cite{jabbari} \cite{daniela}
\cite{zumino}
 \be\label{expo}
 \phi(\hat{\sigma}_{i},\hat{\sigma}_{j})=\sum_{k,p=-\infty}^{\infty}c_{kp}\hat{U}^{k}\hat{U}^{p}
 \ee
For the commutative case, using the symbols corresponding to the $\hat{U_i}$ operators, we can
write an equivalent expression to the above
 \be \label{expr}
\phi(\sigma_{i},\sigma_{j})=
\sum_{k,p=-\infty}^{\infty}c_{kp}~e^{i(\frac{k}{\rho}X_{1}+\frac{p}{\rho}X_{2})}
\qquad X_{i}\in \mathds{R}
 \ee
 In the rational (fuzzy) case, we can
take $\theta = \frac{1}{N}$ without losing generality, where $N$ is
the (finite) size of the representation and the theory on the fuzzy
torus is equivalent to a lattice theory with $X\rightarrow X_{n}=
\frac{2\pi\rho}{N}n$.
 Eq.~(\ref{expr}) becomes \cite{jabbari}
\be\label{expq}
 \phi(n,m)=\sum_{k,p=0}^{N-1}d_{kp}~e^{i\frac{2\pi}{N}(kn+pm)}.
 \ee
We will use the equation above to expand the reduced Green's function
(see below) $g(y,y', k^{\lambda}k_{\lambda})$ on the fuzzy torus.

\section{Green's Function technique}

To obtain the energy density on $\mathcal{M}^4 \times \mathbf{T}^2$,
we first calculate the energy-momentum tensor of a massless scalar
field defined by

 \be
  t_{AB}=\partial_{A}\varphi\partial_{B}\varphi-\frac{1}{2}g_{AB}\partial_{C}\varphi\partial^{C}\varphi\qquad
  (A,B=0\dots 5) \; .
   \ee
In terms of the Green's function the energy density on
$\mathcal{M}^4\times \mathbf{T}^2$ can be written \cite{milton1} as
 \be
 \label{u1} u(\rho)=-\frac{iV_{T^{2}}}{2(2\pi)^4}\int d \vec{k} \int_{c}dw\, w^2 g(y,y,k^{\lambda}k_{\lambda})
 \ee
where $g(y,y,k^{\lambda}k_{\lambda})$ is the reduced Green's
function defined on the extra dimensions $(x_{\mu} \in
\mathcal{M}^4$ and $y=y_{1},y_{2}$ are the two coordinates for the
torus). To find the expression for $g(y,y,k^{\lambda}k_{\lambda})$
we solve the equation of motion satisfied by
$g(y,y',k^{\lambda}k_{\lambda})$. The equations of motion for a
commutative torus differ from that of $\mathbf{T}_{F}^{2}$. We
first perform  the calculations for the commutative case.

\subsection{Green's function for the commutative torus }
Our goal is to calculate the contribution to the vacuum energy
density from the part of the manifold which is a noncommutative
strip with opposite sides identified, which is topologically a
torus. To do this we first calculate the periodic Green's function
for the commutative torus.  We were unable to find the explicit
form for such a Green's function in the literature, and so we
include this calculation for completeness.  We will then adapt
this calculation to the fuzzy torus in the next section.
\par
 When $y_{1}$ and $y_{2}$ are continuous variables, the equation of
motion is the usual Klein-Gordon equation with
$k^{\lambda}k_{\lambda}$ representing the Kaluza-Klein mass term
and $\nabla^{2}_{T^{2}}$ the Laplacian operator defined on a
torus, i.e the ordinary Laplacian defined on $\mathds{R}^2$ but
acting on a peridic function \be\label{gdef}
(\nabla^{2}_{T^{2}}+k^{\mu}k_{\mu})g(y,y',k^{\lambda}k_{\lambda})=-\delta_{P}^2(y-y')\;
, \ee
 with $\delta_{P}$ being a ``periodic'' delta function (see
 Appendix).
To find $g(y,y',k^{\lambda}k_{\lambda}$) we expand it
on the basis $U_{i}$ and specify $y\rightarrow (y_{1},y_{2}) \in
T^{2}$: \be\label{gexp}
 g(y,y',k^{\lambda}k_{\lambda})  = \sum_{k,p=-\infty}^{\infty}
 c_{kp}e^{i(\frac{k}{\rho}y^{}_{1}+\frac{p}{\rho}y^{}_{2})}e^{-i(\frac{k}{\rho}y'_{1}+\frac{p}{\rho}y'_{2})}.
\ee
In order to find $c_{kp}$, we susbstitute Eq.(\ref{gexp}) into
Eq.(\ref{gdef}) and with $ \nabla^{2}_{T^{2}}
U_{i}=-\frac{k^2+p^2}{\rho^2}U_{i}$ we obtain
 \be
\sum_{k,p=-\infty}^{\infty}c_{kp}\Large(
-\frac{k^2+p^2}{\rho^2}+k^{\lambda}k_{\lambda}\Large)e^{i(\frac{k}{\rho}y^{}_{1}+\frac{p}{\rho}y^{}_{2})}e^{-i(\frac{k}{\rho}y'_{1}+\frac{p}{\rho}y'_{2})}
= -\delta_{P}(y^{}_{1}-y'_{1})\delta_{P}(y^{}_{2}-y'_{2})
 \ee
A property of $\delta_{P}(y-y')$ is
  \be
\frac{1}{(2\pi\rho)^2}\sum_{k,p=-\infty}^{\infty}e^{i(\frac{k}{\rho}y^{}_{1}
+\frac{p}{\rho}y^{}_{2})}e^{-i(\frac{k}{\rho}y'_{1}+\frac{p}{\rho}y'_{2})}
=\delta_{P}(y^{}_{1}-y'_{1})\delta_{P}(y^{}_{2}-y'_{2}).
 \ee
  The expression for $c_{kp}$ is therefore
 \be
c_{kp}=\frac{1}{(2\pi\rho)^2(\frac{k^2+p^2}{\rho^2}-k^{\lambda}k_{\lambda})}
\ee We finally obtain for the energy density (GC: Green's function
approach, Commutative case)
 \be\label{u1i}
u_{GC}(\rho)=-\frac{i}{(2\pi)^4}\int d \vec{k} \int_{c_{+}}dw\, w^2
\sum_{k,p=-\infty}^{\infty} \frac{1}{\frac{k^2+p^2}{\rho^2}
+\vec{k}^2-w^2}
 \ee
 It should be obvious from the context that we integrate only over
 the wave vector $\vec{k}$ while summing over the integer $k$.
 To calculate the expression above, we first perform the
integration using the contour argument in
\cite{milton1}.  The $d \vec{k}$ term gives $4\pi k^{2} dk$, then we
perform the change of coordinates $k=R\cos(\theta)$ and
$\omega=iR\sin(\theta)$, and the integral becomes
 \be\label{ugc1}
u_{GC}(\rho)=-\frac{i}{(2\pi)^4}(4\pi)\sum_{k,p=-\infty}^{\infty}\int_{0}^{\infty}
dR ~\int_{-\frac{\pi}{2}}^{\frac{\pi}{2}}d\theta~ \frac{R^5
\cos^{2}(\theta)\sin^{2}(\theta)}{\frac{k^2+p^2}{\rho^2}+R^2 } \; .
 \ee
In order to obtain a finite result we impose a cutoff on the
variable $R$: $R_{max}=\frac{1}{b}$ with $b\simeq L_{Planck}$.
 We also discard the two infinite terms, which are not proportional to $\ln(\frac{\rho}{b})$
 as done previously in \cite{fabi} \cite{milton1}, and we obtain
 \be\label{ugc}
 u_{GC}(\rho)= - \frac{1}{32\pi^{2}\rho^{4}}\sum_{k,p\neq 0,0} \left( {k}^{2}+{p}^{2} \right) ^{2}\ln  \left[1+{\frac {{\rho
  }^{2}}{{b}^{2} \left( {k}^{2}+{p}^{2} \right) }} \right]
.\ee

We first notice that this quantity is always negative and therefore
considering the case of the extra dimensions to be a two dimensional
commutative torus does not provide a valid model for dark energy.
The summation above does not contain the $k=p=0$ term because we
have previously discarded it; it corresponded to the $b^{-4}$
divergence contained in Eq (\ref{ugc1}). Also note that the sum is
even in $k$ and $p$. In order to evaluate the sum present in Eq.
(\ref{ugc}) we introduce a cutoff on the number of modes allowed. We
call this cut off $N$ and introduce the new variables
$x_{k}=\frac{2\pi\rho}{N}k$ and $y_{p}=\frac{2\pi\rho}{N}p$. In the
large $N$ limit these variables become continuous, and the sum can
be approximated by an integral $\sum \rightarrow
\frac{N^2}{(2\pi\rho)^2}  \int$. We obtain \be
 u_{GC}(\rho)=- \frac{1}{32\pi^{2}\rho^{4}}\frac{N^6}{(2\pi\rho)^6}
\int_{0}^{2\pi\rho} dxdy ~
(x^2+y^2)^2\ln\left[1+\Big(\frac{2\pi\rho}{Nb}\Big)^2\frac{\rho^2}{x^2+y^2}\right]\; .
\ee
The integral converges due to the ultraviolet behavior of the
torus factor dominating that of the flat $\mathcal{M}^4$. The regime
in which commutative effective field theory is valid is $x>b$ thus,
respecting this condition for the toroidal theory amounts to
satisfying $\frac{\rho}{b}\sim N$.
The further change of variable $x'=\frac{x}{2\rho}$ allows the integral to be evaluated numerically
with the result
\be \label{umilton}
 u_{GC}(\rho)=-\alpha_{P} \frac{N^6}{32\pi^{2}\rho^4}\; ,
\ee where $\alpha_{P}\simeq 418$
\footnote{Our expression for
Eq.(\ref{umilton}) is the counterpart to the one in Eq.(4.10) of
\cite{jabbari}, which we could not reproduce through numerical
evaluation of the double integral in that equation.}.

\subsection{Green's function for the fuzzy torus}

The assumption that $\theta$ is rational (equal to $\frac{1}{N}$ for
simplicity), implies that the theory becomes equivalent to a lattice
theory with $y\rightarrow y_{n}= \frac{2\pi\rho}{N}n$. The discrete
version of the Laplacian operator used in Eq. (\ref{gdef}) can be
evaluated by performing the variation of the action below (similar to Eq. (4.7) in
\cite{jabbari}) and considering only the two terms relevant to the
torus.
 \be\label{action}
S[\Phi]=\frac{V_{T_{F}^{2}}}{2}\sum_{n,m=0}^{N-1}\int_{\mathcal{M}^{4}}
d^{4}z\left\{(\partial_{\mu}\Phi)^{2}-
\frac{1}{(2\pi\rho\theta)^{2}}(\delta_{n}\Phi)^{2}-
\frac{1}{(2\pi\rho\theta)^{2}}(\delta_{m}\Phi)^{2} \right\}
 \ee
where $\delta_{n}\Phi(z,n,m)\equiv \Phi(z,n+1,m)-\Phi(z,n,m)$ and similarly for $\delta_{m}$.
With the notation: $y_{1}\rightarrow n,~y_{2}\rightarrow m$,
the equation of motion satisfied by the reduced Green's function is
\be\label{gdefr}
\Big(-\frac{N^2}{(2\pi\rho)^2}[(\delta_{n-1}-\delta_{n})+(\delta_{m-1}-\delta_{m})\big]+k^{\lambda}k_{\lambda}\Big)g(n,n',m,m')=
-\frac{N^2}{(2\pi\rho)^2}\delta_{n n'}\delta_{m m'} \ee and the
expansion of the reduced Green's function is
 \be
g(n,n',m,m')=\sum_{k,p=0}^{N-1}d_{kp}~e^{i\frac{2\pi}{N}(kn+pm)}e^{-i\frac{2\pi}{N}(kn'+pm')}
\ee
 The action of the discrete Laplacian on the basis is
$(\delta_{n-1}-\delta_{n})U^{n}=(2-2\cos(\frac{2\pi k}{N}))U^{n}$
and $d_{kp}$ can be found from the relation \be
\sum_{k,p}^{N-1}d_{kp}~\bigg(\frac{-4N^{2}}{(2\pi\rho)^{2}}[\sin^2\frac{\pi
k}{N}+\sin^2\frac{\pi p}{N}] +k^{\lambda}k_{\lambda}\bigg)
e^{i\frac{2\pi}{N}(kn+pm)}e^{-i\frac{2\pi}{N}(kn'+pm')} =
-\frac{N^2}{(2\pi\rho)^2}\delta_{n n'}\delta_{m m'} \ee Using \be
\sum_{k}^{N-1}e^{i\frac{2\pi}{N}k(n-n')}=N\delta_{n
n'}\equiv\delta_{Fnn'} \; , \ee
we find the expression for the coefficient to be
 \be
d_{kp}=\frac{1}{(2\pi\rho)^2\bigg(\frac{4
N^{2}}{(2\pi\rho)^{2}}[\sin^2\frac{\pi k}{N}+\sin^2\frac{\pi
p}{N}]-k^{\lambda}k_{\lambda}\bigg)} \; ,
 \ee
  and we obtain the energy density due to the fuzzy torus to be (GF: Green's function
  approach, Fuzzy torus)
 \be
 u_{GF}(\rho)=-\frac{i}{(2\pi)^4}\int d \vec{k} \int_{c_{+}}dw\, w^2 \sum_{k,p=0}^{N-1}
 \frac{1}{\frac{4N^{2}}{(2\pi\rho)^{2}}[\sin^2\frac{\pi k}{N}+\sin^2\frac{\pi p}{N}]
+\vec{k}^2-w^2}  \; . \ee We calculate the integral in the same way
as in the commutative case. From Eq. (\ref{u1i}) $u_{GF}(\rho) $ is
obtained

\be\label{ugf}
u_{GF}(\rho)={\frac {-N^{4}}{64\pi^{6}\rho^{4}}}\sum_{k,p=0}^{N-1}
 [\sin^2\frac{\pi k}{N}+\sin^2\frac{\pi p}{N}]^{2}
\ln\Big(1+\frac{4\pi^{2}\rho^{2}}{b^{2}N^{2}4[\sin^2\frac{\pi
k}{N}+\sin^2\frac{\pi p}{N}]}\Big) \; .
 \ee
We now simplify the expression above considering the approximation
$1<N\ll \frac{\rho}{b}$  and obtain
 \be\label{ugfa}
 u_{GF}(\rho)\simeq{\frac {-N^{4}}{64\pi^{6}\rho^{4}}}\sum_{k,p=0}^{N-1}
 [\sin^2\frac{\pi k}{N}+\sin^2\frac{\pi p}{N}]^{2}
\ln\Big(\frac{\rho^{2}}{b^{2}N^{2}[\sin^2\frac{\pi
k}{N}+\sin^2\frac{\pi p}{N}]}\Big) \; .
 \ee
Eq. (\ref{ugf}) in the case of maximum noncommuativity ($N=2$) can be expressed as
 \be\label{ugqN2}
u_{GF}(\rho)=-\frac{\alpha_{GF}}{\rho^4}\ln\Big(\frac{\rho}{b}\Big)=-\frac{3}{\pi^{6}}\ln\Big(\frac{\rho}{b}\Big)\simeq
-\frac{0.00312}{\rho^4}\ln\Big(\frac{\rho}{b}\Big) \; .
 \ee

\section{Hamiltonian technique}
In \cite{jabbari} the Casimir energy on $\mathds{R}\times
\mathbf{T}_{F}^{2}$ has been evaluated by expanding the scalar field
$\phi$ into the creation and annihilation operators with the
following result
 \be\label{jab1}
  u(\rho)=\frac{N}{16\pi^{3}\rho^3}\sum_{k,p=0}^{N-1}\sqrt{\sin^{2}\frac{\pi k}{N}+\sin^{2}\frac{\pi p}{N}}.
 \ee
In order to compare with the result given by Eq.(\ref{ugfa}) we
generalize Eq.(\ref{jab1}) to the  $\mathcal{M}^4\times
\mathbf{T}_{F}^{2}$ case. We notice first that the definition of
$\omega$ , Eq.(4.8) in \cite{jabbari}, needs to be modified in order
to include the dependence on the wave vector $\vec{k}$

 \be
\omega_{k,p}^{2}=\frac{N^2}{\pi^{2}\rho^{2}}\left(\sin^{2}\frac{\pi k}{N}+  \sin^{2}\frac{\pi p}{N}\right)+\vec{k}^2
 \ee
The vacuum expectation value multiplied by the volume of the
two-torus $V_{{T}^{2}}=(2\pi\rho)^{2} $ only, gives the energy
density on $\mathcal{M}^4$ (HF: Hamiltonian approach, Fuzzy torus)
\be\label{trho}
 u_{HF}(\rho)=\frac{N \cdot V_{T^2}}{16\pi^{3}\rho^{3}}\sum_{k,p=0}^{N-1}\int \frac{2 d \vec{k}}{(2\pi)^{3}}\sqrt{\left(\sin^{2}\frac{\pi k}{N}+
\sin^{2}\frac{\pi p}{N}\right)
+\frac{\vec{k}^2\pi^{2}\rho^{2}}{N^{2}}}
 \ee
To evaluate (\ref{trho}) we first perform  the integration over $d \vec{k}=4\pi k^2dk$.
 The integral diverges in this case also and needs to be
regularized by 1) introducing a cutoff on $k$ ($k_{max}=\frac{1}{b}$ with $b\simeq L_{Planck}$),
 2) discarding again two infinite terms which are not proportional to $\ln(\frac{\rho}{b})$.
The energy density given by the expansion of the field in terms of
creation and annihilation operators is
 \be\label{uh}
u_{HF}(\rho)=\frac{-N^4}{64\pi^{6}\rho^{4}}\sum_{k,p=0}^{N-1} [
\sin^{2} \frac {\pi k}{N}+\sin^{2}\frac {\pi p}{N}]^{2}
\ln\Bigg(\frac{~\Big(\frac{\rho^{2}\pi}{b}+ \rho\sqrt{[ \sin^{2}
\frac {\pi k}{N}+\sin^{2}\frac {\pi p}{N}
]N^{2}+\frac{\rho^{2}\pi^{2}}{b^{2}}}\quad \Big)^{2}} {[\sin^{2}
\frac {\pi k}{N}+\sin^{2}\frac {\pi p}{N} ]N^{2}\rho^{2} }\Bigg) \; .
 \ee
  Again considering $1<N\ll \frac{\rho}{b}$ we obtain:
 \be\label{uha}
 u_{HF}(\rho)\simeq\frac{-N^4}{64\pi^{6}\rho^{4}}\sum_{k,p=0}^{N-1}
[\sin^{2} \frac {\pi k}{N}+\sin^{2}\frac {\pi p}{N}]^{2}
\ln\Bigg(\frac{\rho^2} {b^2 N^2 [ \sin^{2} \frac {\pi
k}{N}+\sin^{2}\frac {\pi p}{N}] }\Bigg) \; ,
 \ee
 which agrees with Eq.(\ref{ugfa}) obtained using the Green's function
 approach.\\
In the particular case of $N=2$, the energy density given by
Eq.(\ref{uh}) is
 \be\label{uhq}
u_{HF}(\rho)=-\frac{\alpha_{HF}}{\rho^4}\ln\Big(\frac{\rho}{b}\Big)=-\frac{3}{\pi^{6}}\ln\Big(\frac{\rho}{b}\Big)\simeq
-\frac{0.00312}{\rho^4}\ln\Big(\frac{\rho}{b}\Big) \; .
 \ee
Which agrees with the result found in
Eq.(\ref{ugqN2}):~$\alpha_{GF}=\alpha_{HF}$.

\section{Conclusion}

We have explored the Casimir energy density experienced by
observers on $\mathcal{M}^4$ due to extra fuzzy dimensions,
$\mathbf{T}^2_{F}$.  Different approaches yield consistent results
where they are both valid.  We have adhered to general field
theory lore concerning the approach to regularization of the
Casimir energy, extracting the part which depends only
logarithmically on the Planck scale.  However, there are still
some interesting subtleties which we will pursue in a forthcoming
paper.  Chief among these is the proof that $u(\theta)$ is well
behaved as a function of $\theta$. Because of the simplifying
assumption that rational $\theta$ is in the form $\frac{1}{N}$,
the large degree of freedom limit coincides with the commutative
one. More generally $\theta=M/N$. So we can have a large number of
degrees of freedom in the noncommutative theory without going
directly to the commutative limit. This is important since it
would be unpleasant to have the Casimir energy jump by a finite
amount whenever $\theta$ passes from a rational to an irrational
value.  In a theory in which $\theta$ is dynamical, this defect
would zero-out the matrix model contribution (it would be measure
zero).  Thus it remains to show that a sequence of matrix model
energy densities $u(\theta=\frac{M}{N})$ converge to the $\theta$
irrational result, which is in turn Morita equivalent
\cite{zumino} to the commutative torus (which was analyzed here).

\section{Appendix}
The delta function on the torus must respect periodicity. Our
candidate for   $ \delta_{P}(t) $ is given by the $
\lim_{N\rightarrow \infty}  F_{N}(t)$, where  $F_{N}(t)$ is a
periodic function with period $2\pi$ in the $t=\frac{x}{\rho}$
variable \be
 F_{N}(t)=\sum_{k=-N}^{N}
 e^{ikt}=\frac{\sin((2N+1)\frac{t}{2})}{\sin(\frac{t}{2})}
\ee
We prove now the defining properties of a delta function for $t \in [0,2\pi]$.
Considering the following integral of $F_{N}(t)$
 \be
\int_{0}^{2\pi} F_{N}(t) dt =  \int_{0}^{2\pi}
\frac{\sin((N+\frac{1}{2})t)}{\sin(\frac{t}{2})} dt=
\int_{0}^{2\pi(N+\frac{1}{2})}
\frac{\sin(u)}{(N+\frac{1}{2})\sin(\frac{u}{2N+1})} du
 \ee

 with  $u=(N+\frac{1}{2})$t. In the $\lim_{N \rightarrow \infty}$ the integral above of
 $  F_{N}(t)$  becomes
\be
 \lim_{N\rightarrow \infty} \int_{0}^{2\pi} F_{N}(t)
 dt=\frac{2}{\pi}\int_{0}^{\infty}   \frac{\sin(u)}{u}du=1.
 \ee
Now we act on a periodic test function $f(t)$
 \be
\lim_{N\rightarrow\infty}\int_{0}^{2\pi} F_{N}(t) f(t)dt&=&\lim_{N\rightarrow\infty}\int_{0}^{2\pi (N+\frac{1}{2})}
\frac{\sin(u)}{(N+\frac{1}{2})\sin(\frac{u}{2N+1})} f
\left(\frac{u}{N+\frac{1}{2}}\right)du\nonumber \\
&=&  \frac{2}{\pi}\int_{0}^{\infty} f(0)\frac{\sin(u)}{u}du=f(0)
 \ee
where $u$ was defined earlier .
\acknowledgments This research was supported in part by the
University of Alabama's Research Advisory Committee.
\
\end{document}